\documentclass[preprint,groupedaddress,superscriptaddress,amsmath,amssymb,amsthm]{revtex4-1}
\usepackage{subfigure,hyperref,bbm,times}
\usepackage[T1]{fontenc}
\usepackage{braket}
\usepackage{amsthm}
\usepackage{epsfig}
\usepackage{color}
\usepackage{graphicx}
\usepackage{dcolumn}
\usepackage{bm}

\providecommand{\openone}{\leavevmode\hbox{\small1\kern-3.8pt\normalsize1}}

\begin{document}

\title{Non-Markovianity and coherence of a moving qubit inside a leaky cavity}

\author{Ali Mortezapour} 
\email{mortezapour@guilan.ac.ir}
\affiliation{Department of Physics, University of Guilan, P. O. Box 41335--1914, Rasht, Iran}

\author{Mahdi Ahmadi Borji}
\affiliation{Department of Physics, Payam Noor University, P.O. Box 19395-3697, Tehran, Iran}

\author{DaeKil Park}
\affiliation{Department of Electronic Engineering, Kyungnam University, Changwon 631-701, Korea}
\affiliation{Department of Physics, Kyungnam University, Changwon 631-701, Korea}

\author{Rosario Lo Franco}
\email{rosario.lofranco@unipa.it}
\affiliation{Dipartimento di Energia, Ingegneria dell'Informazione e Modelli Matematici, Universit\`{a} di Palermo, Viale delle Scienze, Edificio 9, 90128 Palermo, Italy}
\affiliation{Dipartimento di Fisica e Chimica, Universit\`a di Palermo, via Archirafi 36, 90123 Palermo, Italy}

\date{\today }


\begin{abstract}
Non-Markovian features of a system evolution, stemming from memory effects, may be utilized to transfer, storage, and revive basic quantum properties of the system states. It is well known that an atom qubit undergoes non-Markovian dynamics in high quality cavities. We here consider the qubit-cavity interaction in the case when the qubit is in motion inside a leaky cavity. We show that, owing to the inhibition of the decay rate, the coherence of the traveling qubit remains closer to its initial value as time goes by compared to that of a qubit at rest. We also demonstrate that quantum coherence is  preserved more efficiently for larger qubit velocities. This is true independently of the evolution being Markovian or non-Markovian, albeit the latter condition is more effective at a given value of velocity. We however find that the degree of non-Markovianity is eventually weakened as the qubit velocity increases, despite a better coherence maintenance.
\end{abstract}

\maketitle

\section{Introduction}
\setcounter{equation}{0}
An open quantum system is one in which there is some interaction between the system of interest and its surrounding environment. Generally, the dynamics of open systems is classified into two prime categories: Markovian (memoryless) and non-Markovian (memory-keeping) regime \cite{petru,devegaRMP}. Markovian regime is indicative of the dynamics in which the past states of the quantum system are irrelevant for what would happen in the future, as paradigmatically evidenced by the Gorini-Kossakowski-Sudarshan-Lindblad (GKSL) equation \cite{GKSequation,Lindblad1976}. Such a feature leads to an irreversible flow of the information from the system to the environment, that explains why the initial coherence of a single qubit system exponentially decays in Markovian regime \cite{petru,devegaRMP}. In contrast, non-Markovian regime indicates that the past history of the system does affect the present one \cite{devegaRMP,rivasreview,piiloPRA,RevModPhys.86.1203}. Owing to this nature, information or energy flows back from the environment to the quantum system of interest \cite{aolitareview,chiuri2012,Lopez2010PRA,darrigo2014IJQI,walbornPRA,
lofrancoChapter,leggioPRA,fanchiniSciRep,LaineGuoPRL}. An example of this behavior is reabsorption of emitted photons by an atom in a high quality cavity or in a photonic band gap (PBG) medium \cite{Lopez2010PRA,xu2010PRL,scalaPRA,kaerPRL,yabPRL,johnPRA,lambroRPP,
laurapseudo,lofrancoreview,bellomo2008trapping,bellomo2010PhysScrManiscalco,bellomo2009ASL,
piiloJumps}. Non-Markovian dynamics is also capable to enable information backflows from the environment to the system when the environment is classical \cite{lofrancoChapter,leggioPRA,LoFrancoNatCom,orieux2015,lofrancospinecho,
lofranco2012PRA,mannone2012,darrigo2013hidden,darrigo2012AOP,
lofranco2012PhysScripta,bellomo2012noisylaser,bordone2012,rossiparis,benedetti2013,
trapani2015,zhou2010QIP,wilsonPRB}. Due to the importance of understanding and characterizing non-Markovian quantum evolution, in the recent years great efforts have been devoted to introduce measures to quantify the non-Markovianity of open quantum systems \cite{breuer2009PRL,rivas2010PRL,lorenzoPRA,rivasreview,breuerRMP,bylicka2014,
luoPRA,wolfPRL,luPRA,sabrinaPRL,addisPRAnonMark}. 

One of the first important consequences of memory effects in non-Markovian regime is that some characteristics of the system, such as quantum coherence and entanglement, partially revive during the time evolution \cite{bellomo2007PRL,bellomo2008PRA,devegaRMP,rivasreview,aolitareview,
lofrancoreview,Man:15,carlosIJQI}. These revivals, albeit prolonging the utilization time of the quantum resources, however eventually decay. Nonetheless, it is well known that implementation of quantum computers requires qubits with long-lived coherence. This aim has led to substantial amount of literature attempting to find strategies in order to controlling and protecting coherence and quantum correlations in systems of qubits under different environmental conditions
\cite{PhysRevLett.79.1953, violaPRA,protopopescuJPA,facchiPRA,scalaPRA,Branderhorst638,jing2013,
lofrancoManPRA,lofrancoManSciRep,manPRA2014,Man:15,yanPLA,yangPRL,behzadi,
pezzuttoarXiv,WangarXiv,addisPRA2014,lofrancoQIP,lofrancoPRB,lofrancospinecho,
xuePRA,Zeno1,hartmannNJP,brunoNJP,brunoEPL,matteoEPJD,Man2,non-Mar2,non-Mar3,universalfreezing,aaronson2013PRA,aaronson2013NJP,gerardofrozen,silvaPRL}. 

Within this context, some studies have been recently performed which consider moving atoms interacting with the electromagnetic radiation, in both uniform and accelerated motion \cite{PhysRevA.95.043824,felicettiSciRep,felicettiPRB,PhysRevD.95.025020,PhysRevA.94.032337,MortezapourLPL}, including the case of relativistic velocities simulated by circuit QED setups \cite{felicettiSciRep,felicettiPRB}. In particular, it has been shown \cite{MortezapourLPL} that separated qubits, each in a uniform non-relativistic motion inside a leaky cavity, exhibit the property to preserve their initial entanglement longer than the case of qubits at rest. This result motivates the study of the basic case of a single moving qubit inside a cavity in order to characterize the underlying dynamical mechanisms of such a system. In this paper we address this issue by analyzing in particular the effect of the motion of the qubit on the degree of non-Markovianity and on the dynamics of its coherence. The qubit interacts with the cavity modes during motion. We discuss how one can manipulate non-Markovianity and protect initial coherence of the system by adjusting the velocity of the qubit.

 The paper is organized as follows. In Sec. 2, we give a Hamiltonian description of the dynamics of the considered system and we also determine a state evolution referring to this Hamiltonian. Coherence quantifier of the system is introduced in Sec. 3. In Sec. 4, we briefly discuss about the measure that is going to be utilized to quantify non-Markovianity of the system. In Sec. 5, we present the results of our numerical simulations illustrating the excellent performance of qubit motion in protecting coherence and manipulating non-Markovianity of the system. Finally, Sec. 6 concludes this paper.

\section{Model and state vector evolution}
\label{2-PARTIES} \setcounter{equation}{0}

\begin{figure}[h]
  \centerline{\includegraphics[width=0.45\textwidth,keepaspectratio=true]{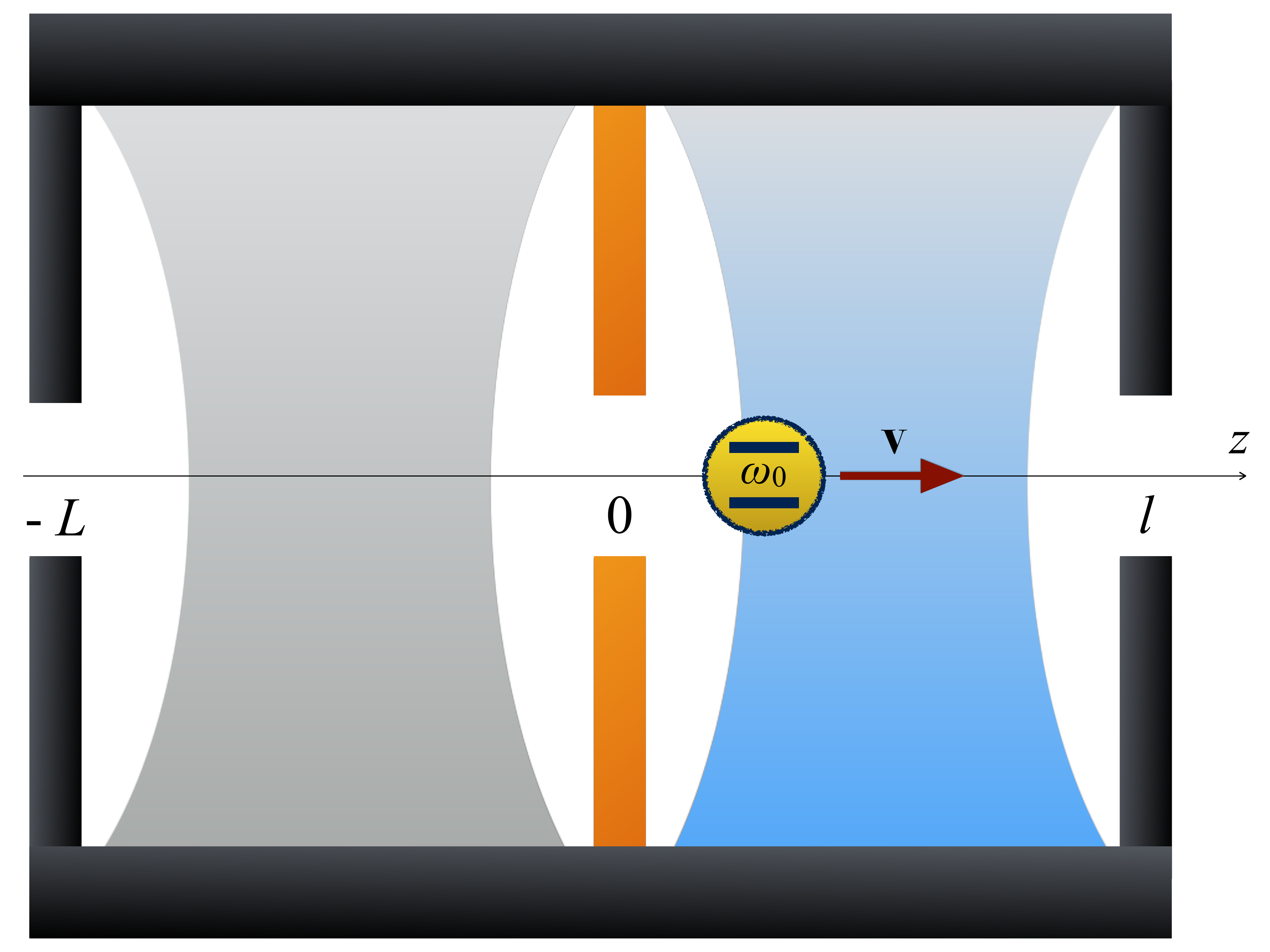}}
  \caption{Schematic illustration of a setup where a single qubit is moving inside a cavity. The qubit is a two-level atom with transition frequency $\omega_0$ traveling with constant velocity \textbf{v}.}\label{fig1}
\end{figure}

We consier a system composed of an atom qubit and a structured environment made of two perfect reflecting mirrors at the positions $z=-L$ and $z=l$ with a partially reflecting mirror in $z=0$. This creates a sort of two consecutive cavities ($-L$, $0$) and ($0$, $l$) as shown in Fig. 1. Any classical electromagnetic field in the system ($-L$, $l$) may be expanded in terms of the exact monochromatic modes $U_{k} (z)$ at frequency $\omega _{k} =ck$ \cite{LangPRA,scullyPRA,leonardi}

\begin{equation} \label{GrindEQ__1_}
E(z,t)=\sum _{k}E_{k} (t)U_{k} (z) + \mathrm{c.c.} ,
\end{equation}
where $E_{k} (t)$ is the amplitude in the $k$-th mode. We assume that the electromagnetic field inside the cavity is polarized along the $\hat{x}$-direction. In order to satisfy the boundary conditions at the mirrors, the mode functions $U_{k} (z)$ must be
\begin{equation} \label{GrindEQ__2_}
U_{k} (z)=\left\{\begin{array}{l} {\xi _{k} \sin k(z+L),\begin{array}{cc} {} & {z<0} \end{array}} \\ {M_{k} \sin k(z-l),\begin{array}{cc} {} & {z>0} \end{array}} \end{array}\right.    ,
\end{equation}
Here $\xi _{k} $ takes the values 1, -1 going from each mode to the subsequent one while $M_{k} $ for a good cavity has the expression \cite{MortezapourLPL,leonardi}
\begin{equation} \label{GrindEQ__3_}
M_{k} =\frac{(c\lambda ^{2} /l)^{1/2} }{[(\omega _{k} -\omega _{n} )^{2} +\lambda ^{2} ]^{1/2} } ,
\end{equation}
where $\omega _{n} =n\pi c/l$  ($n>>1$) are the frequencies of the quasi modes and $\lambda $ is the damping of the ($0$, $l$) cavity. In fact, $\lambda $ quantifies photons leakage through cavity mirrors and indicates the spectral width of the coupling: the larger the quality factor $Q$ of the cavity, the smaller the spectral width $\lambda $.

 The qubit (two-level atom) is taken to interact only with the second cavity ($0$, $l$) and it moves along the z-axis with constant velocity $v$ (see Fig. \ref{fig1}). This condition can be thought to be fulfilled by Stark shifting (for instance, by turning on a suitable external electric field) the atom frequency far out of resonance from the cavity modes until $z=0$ and then turning off the Stark shift \cite{raimondRMP,MortezapourLPL}. During the translational motion, the qubit interacts with the cavity modes.

Under the dipole and rotating-wave approximation, the Hamiltonian of the system reads as (hereafter we take $\hbar \equiv 1$)
\begin{equation} \label{GrindEQ__4_}
\hat{H}=\omega _{0} {\left| a \right\rangle} {\left\langle a \right|} +\sum _{k}\omega _{k} a_{k}^{{\rm \dag }} a_{k}  +\sum _{k}f_{k} (z)[g_{k} {\left| a \right\rangle} {\left\langle b \right|} a_{k} +g_{k}^{*} {\left| b \right\rangle} {\left\langle a \right|} a_{k}^{{\rm \dag }} ]
\end{equation}
where ${\left| a \right\rangle}$ (${\left| b \right\rangle} $) and $\omega _{0} $ are the excited (ground) state and the transition frequency of the qubit. $a_{k}^\dag$ ($a_{k}$) is the creation (annihilation) operator for the $k$-th cavity mode with frequency 
$\omega _{k} $ and $g_{k} =-d(\omega _{k} /\hbar \varepsilon _{0} Al)^{1/2} $ denotes the coupling constant between the qubit and the cavity modes. Notice that $d$ is the magnitude of electric-dipole moment of the atom qubit and $A$ is the surface area of the cavity mirrors.

 The parameter $f_{k} (z)$ describes the shape function of qubit motion along the z-axis, and it is given by \cite{leonardi,MortezapourLPL}
\begin{equation} \label{GrindEQ__5_}
f_{k} (z)=f_{k} (vt)=\sin [k(z-l)]=\sin [\omega _{k} (\beta t-\tau )],
\end{equation}
where $\beta =v/c$ and $\tau =l/c$. Note that the coupling function is not zero when $z=0$, while it is zero when $z=l$ (perfect boundary).

As we mentioned in our previous work \cite{MortezapourLPL}, the translational motion of an atom can be treated classically ($z=vt$) if the de Broglie wavelength $\lambda _{B} $ of the atom is much smaller than the wavelength $\lambda _{0} $ of the resonant transition ($\lambda _{B} /\lambda _{0} <<1$) \cite{MortezapourLPL,cookPRA}. On the other hand, the relative smallness of photon momentum ($\hbar \omega _{0} /c$) compared to atomic momentum ($mv$) allows one to neglect the atomic recoil resulting from the interaction with the electric field \cite{meystrePRA}.  In order to fulfill these conditions, the velocity of the ${}^{85}$Rb Rydberg microwave qubit ($\omega _{0} =51.1$ GHz and $\gamma =33.3$ Hz) \cite{harochePRA,raimondRMP} and an optical qubit ($\omega _{0} \approx 500$ THz and $\gamma \approx 10^{8}$ Hz) are required to be $v>>10^{-7} $ m/s and $v>>10^{-3} $ m/s, respectively. In our following analysis, we shall take into account values of the parameter $\beta$ written as $\beta =(x)\times 10^{-9} $ that, for the ${}^{85}$Rb Rydberg microwave qubit, are equivalent to $v\approx 0.3(x)$ m/s.

 We assume the overall system is initially in a product state with the qubit in a coherent superposition of its basis states $\alpha {\left| a \right\rangle} +\beta {\left| b \right\rangle} $, with $\left|\alpha \right|^{2} +\left|\beta \right|^{2} =1$, and the cavity modes in the vacuum state ${\left| 0 \right\rangle} $, that is
\begin{equation} \label{GrindEQ__6_}
{\left| \Psi (0) \right\rangle} =\{ \alpha {\left| a \right\rangle} +\beta {\left| b \right\rangle} \} {\left| 0 \right\rangle} .
\end{equation}
Hence, at any later time $t$, the overall quantum state can be written as
\begin{equation} \label{GrindEQ__7_}
{\left| \Psi (t) \right\rangle} =\alpha A(t){\left| a \right\rangle} {\left| 0 \right\rangle} +\beta {\left| b \right\rangle} {\left| 0 \right\rangle} +\sum _{k}B_{k} (t){\left| b \right\rangle} {\left| 1_{k}  \right\rangle}  ,
\end{equation}
where ${\left| 1_{k}  \right\rangle} $ represents the cavity field state containing one photon in the $k$-th mode. From the Schr\"{o}dinger equation, one then obtains the differential equations of the probability amplitudes $A(t)$ and $B_{k} (t)$ as
\begin{equation} \label{GrindEQ__8_}
i\dot{A}(t)=\omega _{0} A(t)+\sum _{k}g_{k} M_{k} f_{k} (vt)B_{k} (t) ,
\end{equation}
\begin{equation} \label{GrindEQ__9_}
i\dot{B}_{k} (t)=\omega _{k} B_{k} (t)+g_{k}^{*} M_{k} f_{k} (vt)A(t),
\end{equation}
Solving Eq. \ref{GrindEQ__9_} formally and substituting the solution into Eq. \ref{GrindEQ__8_}, one obtains
\begin{equation} \label{GrindEQ__10_}
\dot{A}(t)+i\omega _{0} A(t)=-\int _{0}^{t}dt^{'} A(t^{'} )\sum _{k}\left|g_{k} \right|^{2} M_{k}^{2} f_{k} (vt)f_{k} (vt^{'} )e^{-i\omega _{k} (t-t^{'} )}   ,
\end{equation}
By redefining the probability amplitude as $A(t^{'} )=\tilde{A}(t^{'} )e^{-i\omega _{0} t^{'} } $, we can rewrite Eq. \ref{GrindEQ__10_} as
\begin{equation} \label{GrindEQ__11_}
\dot{\tilde{A}}(t)+\int _{0}^{t}dt^{'}  F(t,t^{'} )\tilde{A}(t^{'} )=0.
\end{equation}
where the kernel $F(t,t^{'} )$, which is the correlation function including the memory effect, has the following form
\begin{equation} \label{GrindEQ__12_}
F(t,t^{'} )=\sum _{k}\left|g_{k} \right|^{2} M_{k}^{2} f_{k} (vt)f_{k} (vt^{'} )e^{-i(\omega _{k} -\omega _{0} )(t-t^{'} )}  .
\end{equation}
This function in the continuum limit becomes
\begin{equation} \label{GrindEQ__13_}
F(t,t^{'} )=\int _{0}^{\infty }J(\omega _{k} )\sin [\omega _{k} (\beta t-\tau )]\sin [\omega _{k} (\beta t^{'} -\tau )]e^{-i(\omega _{k} -\omega _{0} )(t-t^{'} )}  d\omega _{k} ,
\end{equation}
with $J(\omega _{k} )$ as the spectral density of an electromagnetic field inside the cavity. Assuming that the employed cavities are imperfect, $J(\omega _{k} )$ gets the following Lorentzian shape \cite{petru}
\begin{equation} \label{GrindEQ__15_}
J(\omega _{k} )=\frac{1}{2\pi } \frac{\gamma \lambda ^{2} }{[(\omega _{0} -\omega _{k} -\Delta )^{2} +\lambda ^{2} ]} ,
\end{equation}
where $\Delta =\omega _{0} -\omega _{n} $ is the detuning between $\omega _{0} $ and the center frequency of the cavity modes ($\omega _{n} $). $\gamma =(d^{2} \omega _{n} /\hbar \varepsilon _{0} Al)$ is the decay rate of the qubit in the Markovian limit of flat spectrum when the qubit is at rest in a position in which $U_{k} (z)$ has a maximum. The relaxation time scale $\tau _{q} $ over which the state of the system changes is related to $\gamma $ by $\tau _{q} \approx \gamma ^{-1} $. As noted above, the parameter $\lambda $ indicates the spectral width of the coupling and it is related to the cavity correlation time $\tau _{cav} $ via $\tau _{cav} =\lambda ^{-1} $.

The explicit expression of the time-dependent coefficient $A(t)$ can be analytically obtained as reported in Ref. \cite{parkarXiv}. In the continuum limit ($\tau \to \infty $) and when $t>t'$, analytic solution of Eq. \ref{GrindEQ__13_} gives rise to
\begin{equation} \label{GrindEQ__16_}
F(t,t^{'} )=\frac{\gamma \lambda }{4} \cosh [\theta (t-t')]e^{-\bar{\lambda }(t-t')}
\end{equation}
where $\bar{\lambda }=\lambda -i\Delta $ and $\theta =\beta (\bar{\lambda }+i\omega _{0} )$. Inserting Eq. \ref{GrindEQ__16_} into Eq. \ref{GrindEQ__11_} and solving the resultant equation, using Bromwich integral formula, one has $\tilde{A}(t)$ as
\begin{equation} \label{GrindEQ__17_}
\tilde{A}(t)=\frac{(q_{1} +u_{+} )(q_{1} +u_{-} )}{(q_{1} -q_{2} )(q_{1} -q_{3} )} e^{q_{1} \gamma t} -\frac{(q_{2} +u_{+} )(q_{2} +u_{-} )}{(q_{1} -q_{2} )(q_{2} -q_{3} )} e^{q_{2} \gamma t} +\frac{(q_{3} +u_{+} )(q_{3} +u_{-} )}{(q_{1} -q_{3} )(q_{2} -q_{3} )} e^{q_{3} \gamma t},
\end{equation}
where the quantities $q_{i}$ ($i=1,2,3$) are the solutions of the cubic equation
\begin{equation} \label{GrindEQ__18_}
q^{3} +2(y_{1} -iy_{3} )q^{2} +(u_{+} u_{-} +y_{1} /4)q+y_{1} (y_{1} -iy_{3} )/4=0,
\end{equation}
with $y_{1} =\lambda /\gamma $, $y_{2} =\omega _{0} /\gamma $, $y_{3} =\Delta /\gamma $ and $u_{\pm } =(1\pm \beta )y_{1} \pm i\beta y_{2} -i(1\pm \beta )y_{3} $.

The evolved density matrix of the qubit in the basis $\{| a \rangle ,| b \rangle \}$ is
\begin{equation} \label{GrindEQ__19_}
\rho (t)=\left(\begin{array}{cc} {\left|\alpha \right|^{2} \left|A(t)\right|^{2} } & {\alpha \beta ^{*} A(t)} \\ {\alpha ^{*} \beta A^{*} (t)} & {1-\left|\alpha \right|^{2} \left|A(t)\right|^{2} } \end{array}\right).
\end{equation}
Taking the derivative of Eq. \ref{GrindEQ__19_} with respect to time, we get
\begin{equation} \label{GrindEQ__20_}
\dot{\rho }(t)=-i\frac{\Omega (t)}{2} [\sigma _{+} \sigma _{-} ,\rho (t)]+\frac{\Gamma (t)}{2} [2\sigma _{-} \rho (t)\sigma _{+}-\sigma _{+} \sigma _{-} \rho (t)-\rho (t)\sigma _{+} \sigma _{-} ],
\end{equation}
where $\Omega (t)=-2\mathrm{Im}\left[\frac{\dot{A}(t)}{A(t)} \right]$ and $\Gamma (t)=-2\mathrm{Re}\left[\frac{\dot{A}(t)}{A(t)} \right]$. The quantity $\Omega (t)$ plays the role of a time-dependent Lamb shift and $\Gamma (t)$ can be interpreted as a time-dependent decay rate \cite{petru}.

\section{Coherence}
\label{MULTI} \setcounter{equation}{0}

Due to the role played by quantum coherence as a resource for quantum technologies, many proposals for its quantification has been advanced \cite{streltsovReview}. An intuitive quantification of quantum coherence is based on the off-diagonal elements of the desired quantum state, being these related to the basic property of quantum interference. Indeed, it has been recently shown \cite{plenioCoherence} that the functional
\begin{equation} \label{GrindEQ__21_}
C(t)=\sum _{i,j(i\ne j)}\left|\rho _{ij} (t)\right| ,
\end{equation}
where $\rho _{ij} (t)$ ($i\ne j$) are the off-diagonal elements of the system density matrix, satisfies the physical requirements which make it a proper coherence measure. Assuming $\alpha =\beta =1/\sqrt{2} $, that gives a maximal initial coherence $C(0)=1$, from Eq. (\ref{GrindEQ__19_}) one immediately finds that the qubit coherence at time $t$ is $C(t)=\left|A(t)\right|$.

\section{Non-Markovianity}
In order to quantify and discuss the non-Markovian behavior in our system, among the various quantifiers, we employ the BLP measure \cite{breuer2009PRL}, based on the trace distance between two quantum states. For a given pair of initial states $\rho _{1} (0)$ and $\rho _{2} (0)$ of the system, the change of the dynamical trace distance is described by its time derivative
\begin{equation} \label{GrindEQ__22_}
\sigma [t,\rho _{1} (0),\rho _{2} (0)]=dD[\rho _{1} (t),\rho _{2} (t)]/dt,
\end{equation}
where $\rho _{1} (t)$, $\rho _{2} (t)$ are the dynamical states corresponding to the initial states $\rho _{1} (0)$, $\rho _{2} (0)$ and the trace distance is defined as $D[\rho _{1} (t),\rho _{2} (t)]=(1/2)\mathrm{Tr} \left|\rho _{1} (t)-\rho _{2} (t)\right|$ with $\left|X\right|=\sqrt{X^{{\rm \dag }} X} $ \cite{nielsenchuang}. It is noteworthy that, in quantum information, the trace distance is related to the distinguishability between quantum states, while its time derivative ($\sigma $) means a flow of information between the system and its environment.

According to this measure, Markovian processes satisfy $\sigma \le 0$ for all pairs of initial states $\rho _{1,2} (0)$ at any time $t$. Physically, this means that both the initial states will eventually lose all their initial information into the environmental degrees of freedom and become indistinguishable. However, if there exists a pair of initial states such that $\sigma >0$ for some time intervals, then an information backflow appears from the environment to the system and the process is deemed non-Markovian. Based on these arguments, it is thus possible to define a measure of non-Markovianity as   \cite{breuer2009PRL}
\begin{equation} \label{GrindEQ__23_}
N=\mathop{\max }\limits_{\rho _{1} (0),\rho _{2} (0)} \int _{\sigma >0}\sigma [t,\rho _{1} (0),\rho _{2} (0)]dt,
\end{equation}
where the time integration is extended over all the time intervals in which $\sigma $ is positive, and the maximization is made over all possible pairs of initial states $\rho _{1} (0)$ and $\rho _{2} (0)$.

\begin{figure*}
  \begin{centering}
  \includegraphics[width=6.5in, keepaspectratio=true]{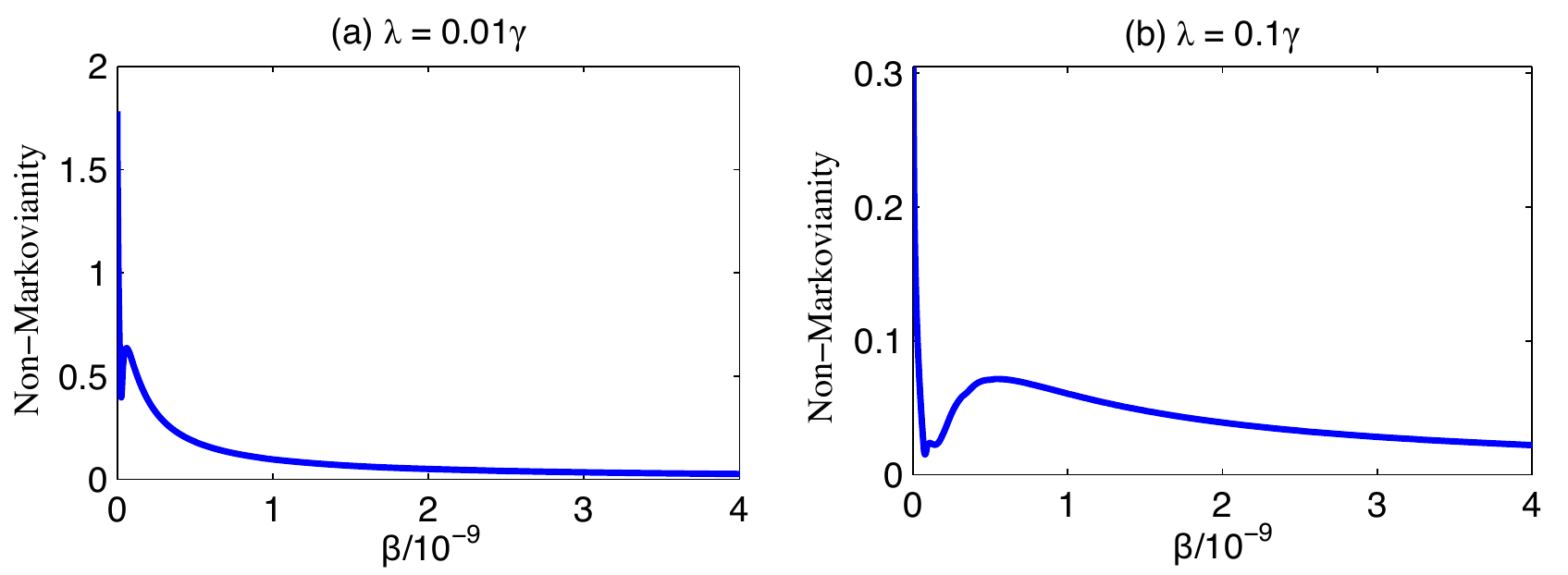}
  \caption{Non-Markovianity as a function of $\beta $ for (a) $\lambda =0.01\gamma$ and (b) $\lambda =0.1\gamma$. Other parameters are taken as: $\alpha =1$, $\Delta =0$, $\omega _{0} =1.53\times 10^{9} $ Hz. }\label{fig2}
  \end{centering}
\end{figure*}

\section{Results and discussion}
In this section, we investigate the effect of qubit velocity on the degree of non-Markovianity and on the time evolution of quantum coherence under various conditions defined by specific values of system parameters.

\begin{figure}
\centerline{\includegraphics[width=3.5in, keepaspectratio=true]{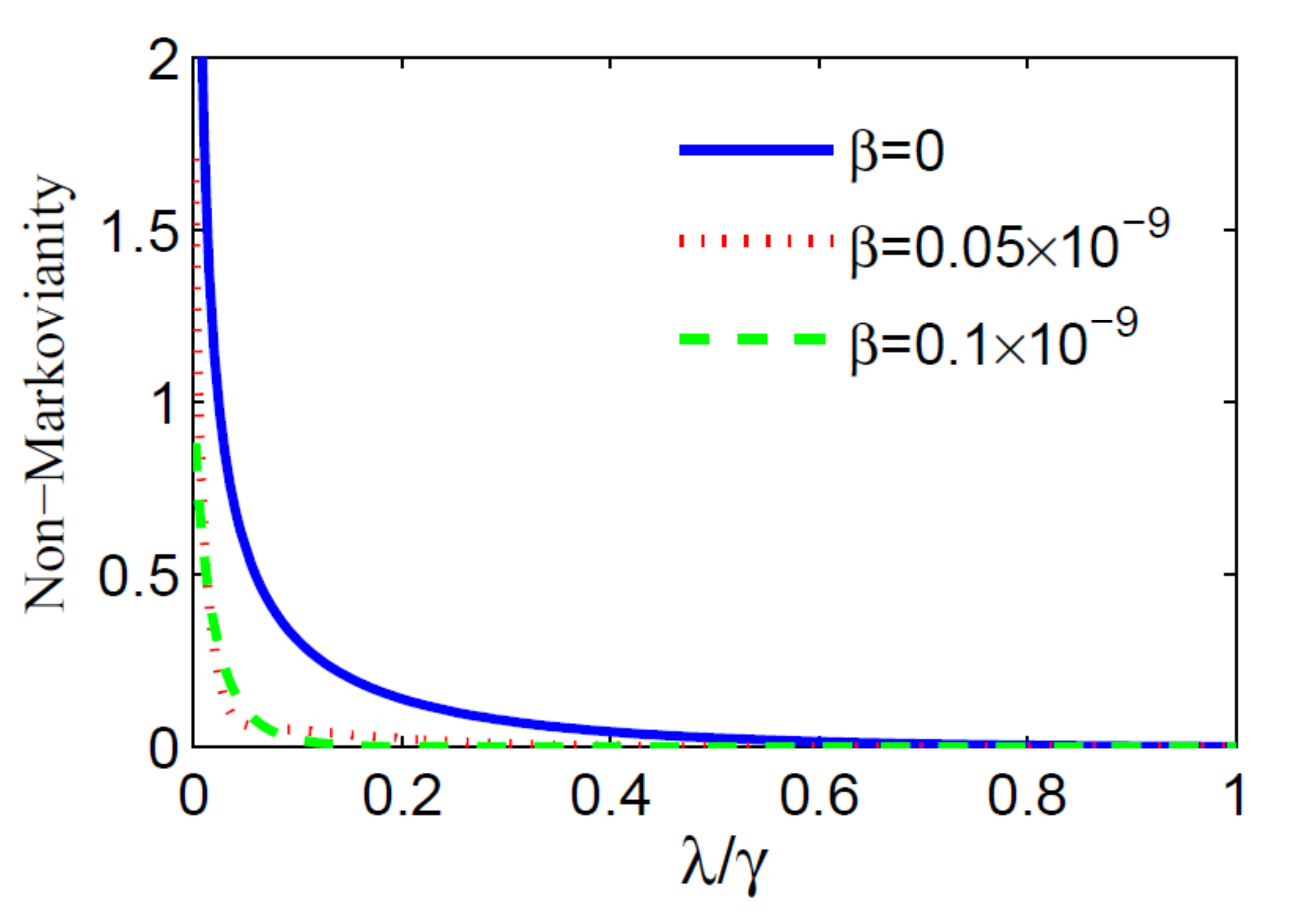}}
  \caption{Non-Markovianity as a function of $\lambda$ for $\beta=0$, $0.05\times10^{-9}$, $0.1\times10^{-9}$. Other parameters are taken as in Fig. 2.}\label{fig3}
\end{figure}

 We plot the degree of non-Markovianity as function of qubit velocity for $\lambda =0.01\gamma$ and $\lambda =0.1\gamma$ in Fig. 2. These values of the spectral bandwidth assures the memory effects of the cavity are prominent \cite{lofrancoreview}. It is seen that non-Markovianity decreases, with some fluctuations, by increasing qubit velocity. Our calculations show that, for a given value of the cavity bandwidth $\lambda$, qubit motion with high enough velocity could completely remove the quantum memory of the system evolution.

\begin{figure*}
\begin{centering}
  \centerline {\includegraphics[width=6.5in, keepaspectratio=true]{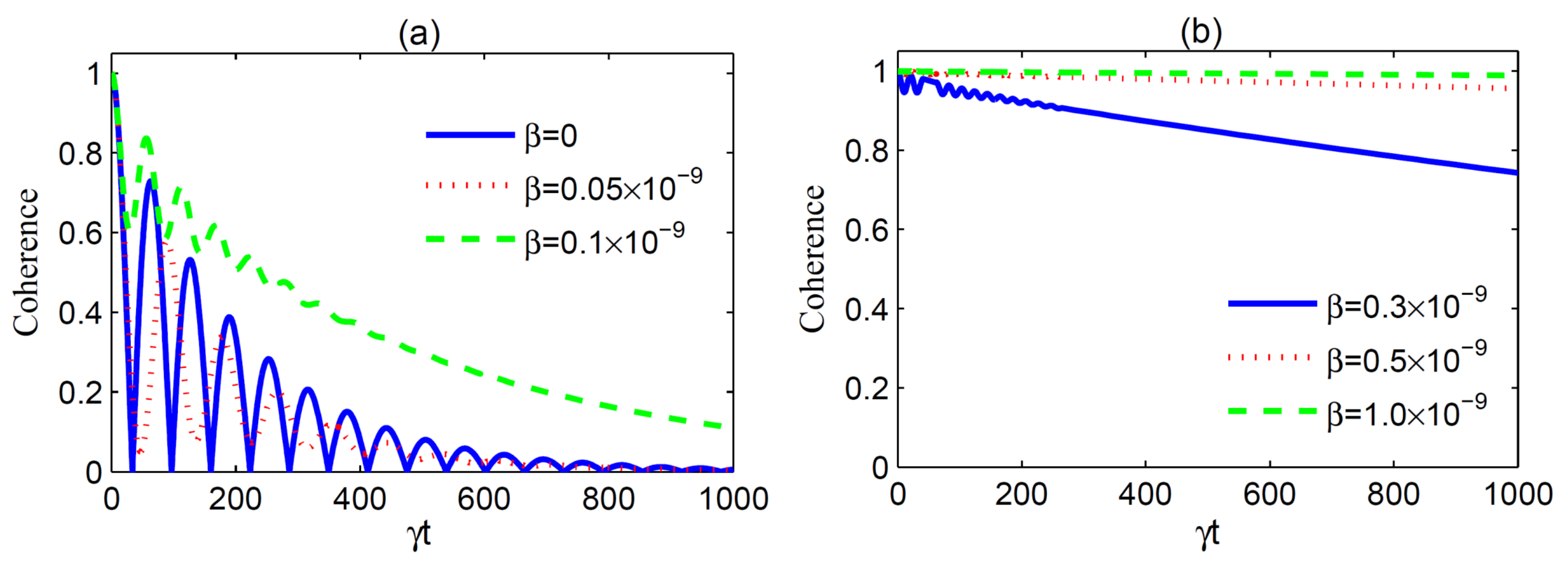}}
  \caption{Coherence $C(t)$ as a function of dimensionless scaled time $\gamma t$ for various velocities of the qubit: (a) $\beta =0$ (solid blue line), $\beta =0.05\times 10^{-9} $ (dotted red line), $\beta =0.1\times 10^{-9} $ (dashed green line); (b) $\beta =0.3\times 10^{-9} $ (solid blue line), $\beta =0.5\times 10^{-9}$ (dotted red line) and $\beta =1.0\times 10^{-9}$ (dashed green line). Values of the other parameters are: $\alpha =\beta =1/\sqrt{2} $, $\lambda =0.01\gamma$, $\Delta =0$, $\omega _{0} =1.53$ GHz.}\label{fig4}
\end{centering}
\end{figure*}

 Now, let us analyze how the qubit velocity influences the range of spectral bandwidth of the cavity ($\lambda$) within which non-Markovian features of the system emerge. Along this route, we plot the degree of non-Markovianity versus $\lambda$ for different velocities of the qubit in Fig. 3. As can be seen, the qubit motion causes a backward shift of the range of values of $\lambda$ for which non-Markovianity is significant. Comparing Fig. 2 and Fig. 3, we can conclude that the qubit motion could remarkably disturb the backflow of information from environment to the qubit. The greater the velocity of the qubit, the smaller the values of $\lambda$ required in order to have a high degree of non-Markovianity.

\begin{figure*}
  \begin{centering}
  \centerline{\includegraphics[width=6.5in, keepaspectratio=true]{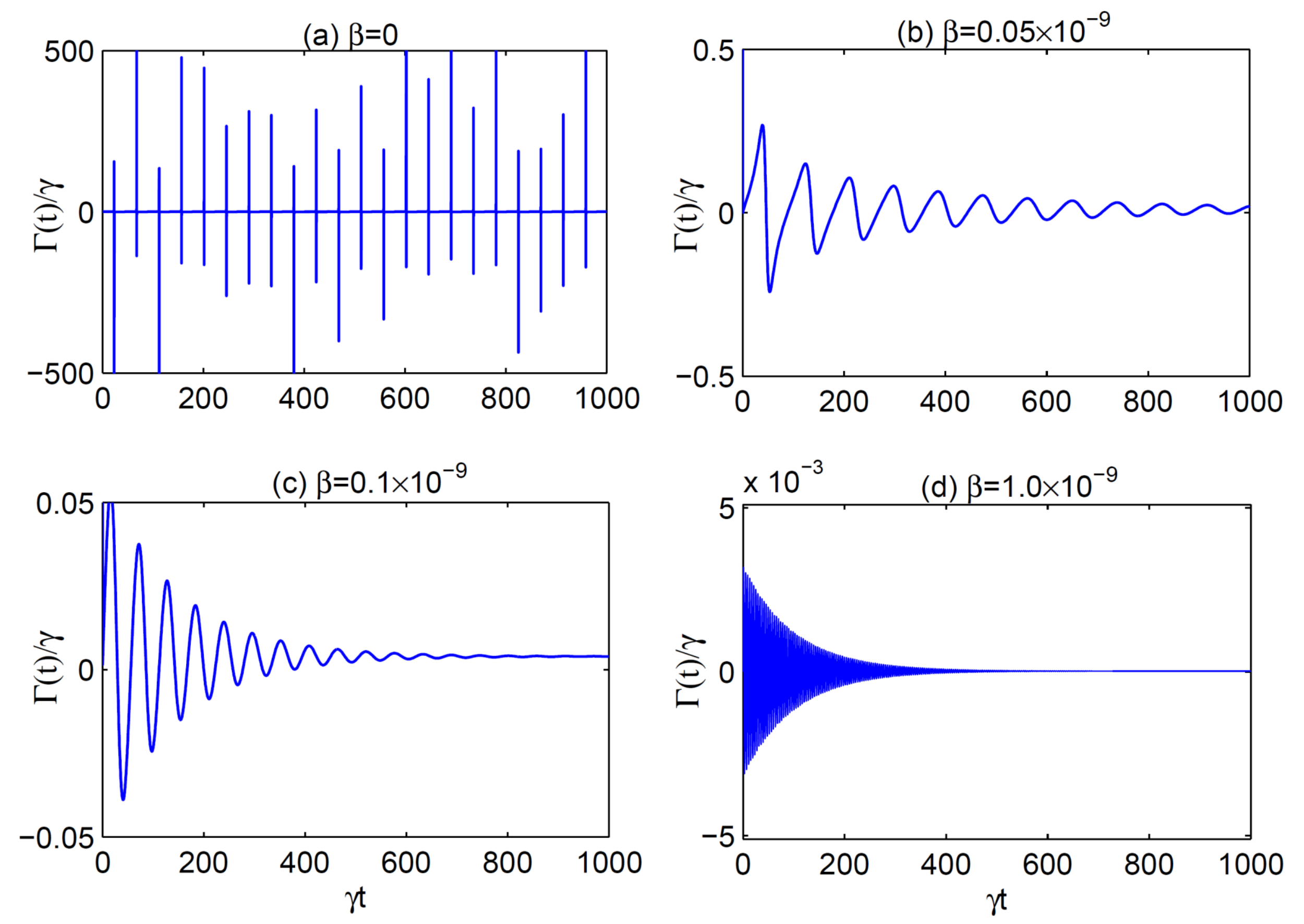}}
  \caption{The decay rate $\Gamma (t)$ as a function of dimensionless scaled time $\gamma t$ for various velocities of the qubit: (a) $\beta =0$; (b) $\beta =0.05\times 10^{-9} $; (c) $\beta =0.1\times 10^{-9} $; (d) $\beta =1.0\times 10^{-9} $. Other parameters are the same as those used in Fig. \ref{fig4}.}\label{fig5}
  \end{centering}
\end{figure*}

 In Fig. 4, we display how the coherence $C(t)$ evolves for various velocities of the qubit under the non-Markovian regime described in Fig. 2(a). It is seen that in the cases of stationary qubit ($\beta =0$), the coherence appears to damp out or collapse and then after a short time, it starts to revive. At longer times, one can find a sequence of collapses and revivals. The amplitudes of the revivals decreases as time goes by. Nevertheless, the qubit coherence settles down and vanishes after some fluctuations in long-time. For slowly moving qubit (e.g., $\beta = 0.05\times 10^{-9} $), the time behavior is similar to the case of stationary qubit with the difference that, during the collapse process, the coherence does not vanish. However, interesting results can be obtained by increasing the velocity of the qubit. In despite of decreasing the oscillating nature (associated to the degree of non-Markovianity), the initial coherence is strongly protected against the noise.

 To better understand the main physical mechanism behind the coherence preservation, we plot in Fig. 5 the decay rate of qubit ($\Gamma (t)/\gamma $) as a function of $\gamma t$ for the velocities of the qubit considered in Fig. 4. The panels of Fig. 5 show how the velocity may affect the decay rate of the qubit. As can be seen, the decay rate oscillates during the evolution, while its (pseudo)period monotonously decreases as qubit velocity increases. Interestingly, there are values of the qubit velocity which can significantly inhibit the decay rate. These plots evidence that the main reason for the efficient preservation of coherence observed above is just the inhibition of the decay rate.

In order to further stress the role of the qubit velocity in reducing the decay rate of the qubit evolution, we finally consider the Markovian regime for the system evolution, choosing a spectral bandwidth which satisfies the weak coupling and memoryless condition, for instance $\lambda = 3\gamma$ \cite{lofrancoreview}. The plots displayed in Fig. 6 give evidence of the fact that increasing the velocity of the atom qubit leads to an enhancing of coherence preservation during the evolution, due to a corresponding decrease of the decay rate as shown in Fig. 7. However, comparing these plots with the previous ones obtained in the non-Markovian regime (see Fig. 4), it is readily seen that velocities much higher are required in the Markovian case than in the non-Markovian case for reaching a comparable level of preservation efficiency. In other words, for a given velocity of the qubit, quantum coherence is better preserved in the presence of non-Markovianity rather than under Markovian (memoryless) conditions.

\begin{figure*}
  \begin{centering}
  \centerline{\includegraphics[width=6.5in, keepaspectratio=true]{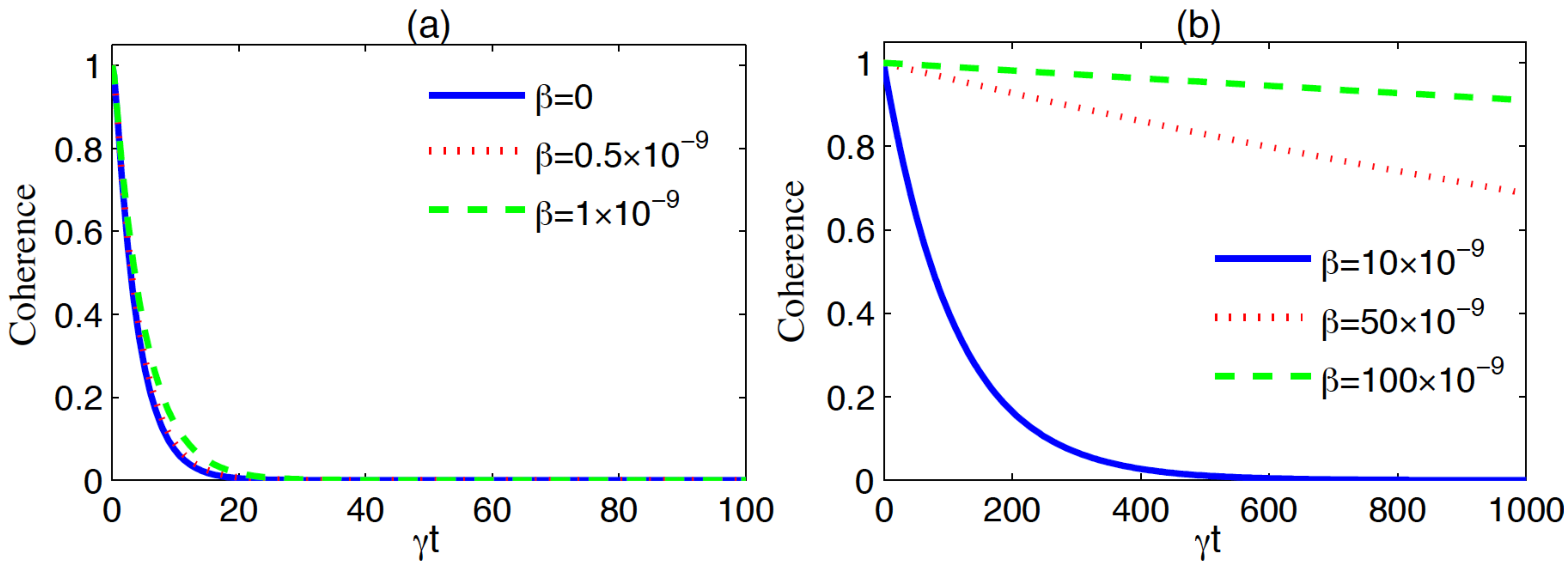}}
  \caption{Coherence $C(t)$ in the Markovian regime as a function of dimensionless scaled time $\gamma t$ for various velocities of the qubit: (a) $\beta =0$ (solid blue line), $\beta =0.5\times 10^{-9} $ (dotted red line), $\beta = 1\times 10^{-9} $ (dashed green line); (b) $\beta = 10\times 10^{-9} $ (solid blue line), $\beta =50 \times 10^{-9}$ (dotted red line) and $\beta =100 \times 10^{-9}$ (dashed green line). Values of the other parameters are: $\lambda = 3\gamma$, $\alpha =\beta =1/\sqrt{2} $, $\Delta =0$, $\omega _{0} =1.53$ GHz.}\label{fig6}
  \end{centering}
\end{figure*}

\begin{figure*}
  \begin{centering}
  \centerline{\includegraphics[width=6.5in, keepaspectratio=true]{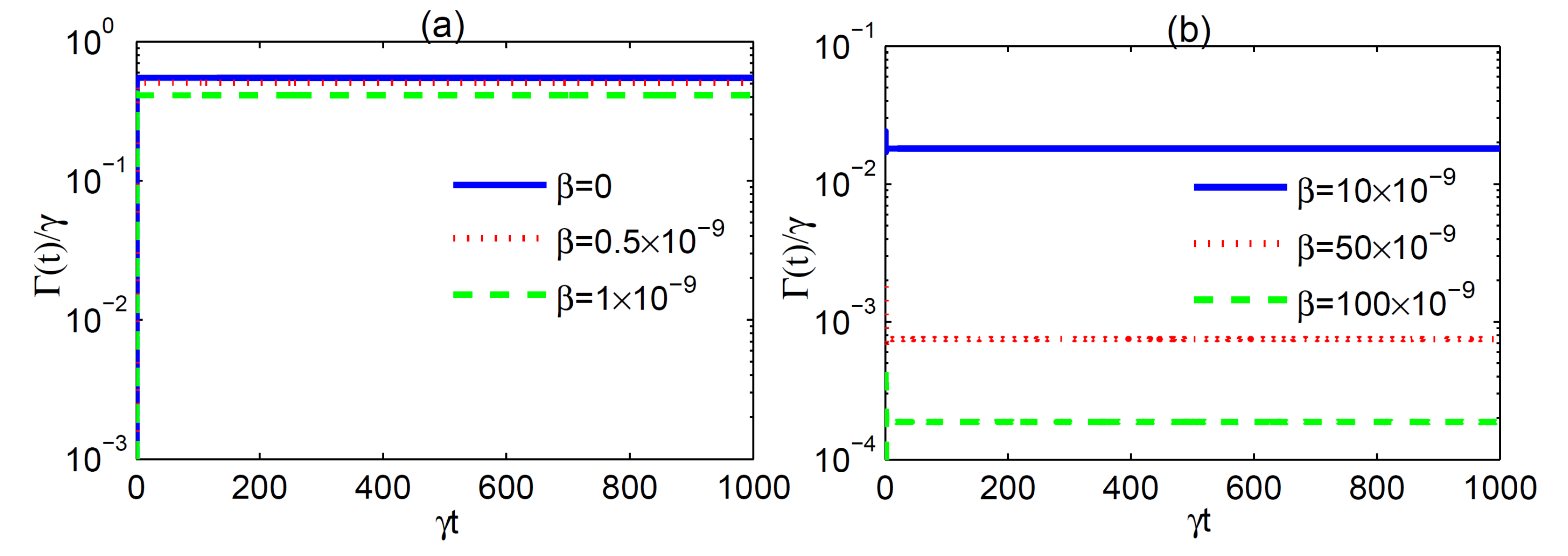}}
  \caption{The decay rate $\Gamma (t)$ in the Markovian regime as a function of dimensionless scaled time $\gamma t$ corresponding to the same values of qubit velocities and parameters of Fig. \ref{fig6}. The plot is in logaritmic scale.}\label{fig7}
  \end{centering}
\end{figure*}

\section{Conclusion}
In this paper we have investigated how qubit motion, inside a leaky cavity, may influence both the quantum coherence initially possessed by an atom qubit and the degree of non-Markovianity of the system evolution. We have found that the motion of the qubit gives rise to a non-monotonic weakening of the non-Markovian features of the system, as identified by information backflows from the environment to the qubit. In particular, increasing the values of the velocity of the qubit, the degree of non-Markovianity eventually tends to decrease. This fact implies that, if one is interested in utilizing non-Markovianity as a resource (for instance, exploiting the oscillations of quantum coherence), the quality factor of the cavity must be improved (that is, very narrow spectral bandwidth) whenever the velocity of the qubit is set to greater values. 

On the other hand, we have shown that the motion of the qubit may act as a protector of  qubit coherence. In fact, the coherence evolves remaining closer to its initial value when the qubit velocity is increased, independently of the evolution being Markovian or non-Markovian. The non-Markovian regime however results more effective than the Markovian one in protecting coherence at a given velocity. We have provided values of the system parameters, namely cavity spectral bandwidth and qubit velocity, which permit a very efficient preservation of quantum coherence. Therefore, for those tasks where quantum coherence constitutes the main resource, our system suggests that the motion of a qubit may enrich the performance during the evolution thanks to a well-preserved coherence. We have clarified that the relevant physical mechanism behind this effective protection of the coherence of the qubit is the inhibition of the decay rate caused by suitable values of the qubit velocity. 
We mention that the model here studied may be also implemented by circuit quantum electrodynamics (cQED) technologies, which are the superconducting analog of the standard cavity-QED setups. Indeed, recent systems of cQED can produce a qubit-cavity coupling strength depending on the qubit position according to a sinusoidal function (like that of Eq. (\ref{GrindEQ__5_})) \cite{jonesSciRep,shanksNatCommun}.

Our results highlight that the qubit motion has an opposite effect on non-Markovianity degree (memory) and quantum coherence. Non-Markovianity and coherence are different resources which can thus behave differently when subject to the same kind of system-environment interaction. Nevertheless, non-Markovianity remains very important for an efficient maintenance of the quantum properties of an open system \cite{lofrancoreview,lofrancoChapter}.

%


\end{document}